\documentclass[journal]{IEEEtran}
\usepackage[dvips]{graphicx}
\usepackage{amssymb}

\begin{document}
\title{Lithium-loaded scintillators coupled to a custom-designed silicon
photomultiplier array for neutron and gamma-ray detection}
\author{Felix~Liang,~\IEEEmembership{Member,~IEEE,}
  Hartmut~Brands,~\IEEEmembership{Member,~IEEE,}
  Les~Hoy,~\IEEEmembership{Member,~IEEE,}
  Jeff~Preston,~\IEEEmembership{Member,~IEEE,}
  and Jason~Smith~\IEEEmembership{Member,~IEEE,}
\thanks{Felix Liang, Hartmut Brands, Les Hoy, and Jason Smith are with FLIR
Systems Inc.}
\thanks{Jeff Preston, formerly with FLIR Systems Inc., is now with Consolidated
Nuclear Security, LLC.}
}
\maketitle

\begin{abstract}
Scintillators capable
of detecting both neutrons and gamma-rays have generated considerable
interest. In particular, the use of such scintillators with
silicon photomultipliers (SiPMs) enables
low-power and compact-geometry applications. Three types of
Li-loaded scintillators, CLYC, CLLB, and NaIL, have been tested with a
custom-designed SiPM array for temperatures between --20 and 50$^{\circ}$C.
The array consists of four 6x6~mm$^2$ SiPMs arranged in a 2x2 configuration.
Pulse shape discrimination is used for neutron and gamma identification.
Because the pulse shape changes with temperature,
the quality of neutron and gamma discrimination varies with temperature.
Furthermore, the larger dark current in SiPMs at high temperatures results in
poorer energy resolution and neutron-gamma discrimination. Comparison of the
energy resolution and the neutron-gamma discrimination for the
three scintillators coupled to the custom SiPM array will be discussed.
\end{abstract}

\section{Introduction}
Helium-3 filled neutron detectors have been widely used in basic science
research, industrial applications, and homeland security.
Since the recent $^3$He shortage, a great deal of effort has been dedicated to
find replacement materials for neutron detection \cite{kou15}. Some Li-loaded
scintillators that are capable of detecting both neutrons and gamma-rays have
generated considerable interest. When
enriched $^6$Li is used in these scintillators, the sensitivity for neutron
detection can be comparable to or exceed $^3$He tubes.

The recent advance in silicon photomultipliers (SiPMs) has made the photo
detection
efficiency similar to the conventional photomultiplier tubes (PMTs).
Due to the compact size, ruggedness, low operating
voltage, and insensitivity to magnetic fields, SiPMs have replaced
PMTs as the photo sensor in some scintillation detectors. However, the active
area of the largest SiPMs is far smaller than that of PMTs. Since the
sensitivity of a radiation detector increases with the detector size,
it is necessary
to assemble SiPMs in an array for larger scintillators. A 2x2 SiPM array has
been designed and constructed for coupling with 18~mm cubic
scintillators \cite{liang16}.
The gain of the SiPMs in the array are matched in the temperature range between
--20 and 50$^{\circ}$C for optimizing the energy resolution. With this SiPM
array coupled to an 18~mm CsI cube, energy resolution of 6.4\% for the 662~keV
gamma-ray has been observed. 

This work investigates the performance of
Cs$_2$LiYCl$_6$:Ce (CLYC) \cite{rmdclyc},
Cs$_2$LiLaBr$_6$:Ce (CLLB) \cite{sgcllb}, and
NaI:Tl,Li (NaIL) \cite{sgnail} coupled
to the custom SiPM array
for neutron and gamma-ray detection. Some physical properties of these
scintillators are given in Table~\ref{tb:scint}.
\begin{table}[!t]
\centering
\caption{Density ($\rho$), wavelength of maximum emission ($\lambda$), light
  yield (ph/MeV), and decay time ($\tau_{d}$) of CLYC, CLLB, and NaIL.}
\begin{tabular}{llll} \hline
    scintillator & CLYC & CLLB & NaIL \\ \hline
    $\rho$ (g/cm$^3$) & 3.3 & 4.2 & 3.6 \\
    $\lambda$ (nm) & 370 & 420 & 418 \\
    ph/MeV & 20,000 & 45,000 & 35,000 \\
    $\tau_{d}$ & 1, 50, 1000 & 180, 1140 & 230, 1000 \\ \hline
\end{tabular}
\label{tb:scint}
\end{table}

\section{Experimental Methods}
Two CLYC scintillators that were manufactured in 2014\footnote{Radiation
  Monitoring Devices, Inc.} and 2016\footnote{CapeSym Inc.}, were used in
this experiment.
The CLYC and CLLB\footnote{Saint-Gobain} scintillators were 18~mm cubes. One
of the surfaces was coupled to the
custom SiPM array where four 6x6~mm$^{2}$ SensL C-Series SiPMs were arranged
in a 2x2
configuration. The rest of the scintillator surfaces were wrapped with
PTFE tape. The scintillator and SiPM array were enclosed in a hermetically
sealed aluminum can to keep out moisture. The detectors were packaged in
a dry glove box in the laboratory.

The NaIL\footnote{Saint-Gobain} scintillator was a 1.5-inch right cylinder
encapsulated in an aluminum
can. The scintillation light was transmitted out of one end of the cylinder by
a quartz window and detected by the SiPM array. Because the 1.5-inch
circular area is considerably larger than the footprint of the SiPM array,
the exposed window area was covered with PTFE tape to reduce light loss. 

The detector output was amplified by a Canberra 2022 spectroscopic amplifier
and recorded by a CAEN V1785N peak-sensing ADC for studying the gamma-ray
spectra. A Struck SIS3302 waveform digitizer was used to record the detector
output for analyzing the pulse shape to discriminate neutrons from
gamma-photons. Because there is a long-decay component of the pulses,
the waveform digitizer stored each trace in a 10 ns interval for 5~$\mu$s.

\section{Results and Discussion}

\subsection{Energy Resolution}
\begin{figure}[!t]
\centering
\includegraphics[width=3.0in]{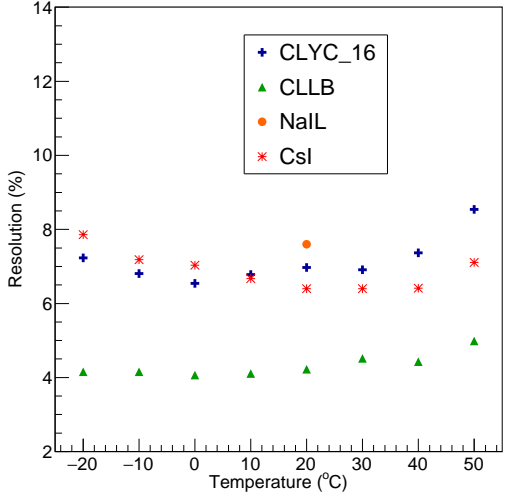}
\caption{Energy resolution for CLYC (crosses), NaIL (circles), and
  CLLB (triangles) as a function of
temperature. CLYC\_16 is the CLYC scintillator manufactured in 2016.
The energy resolution for CsI (asterisks) is shown for comparison.}
\label{fg:resoltemp}
\end{figure}

The energy resolution of the detectors were measured with a $^{137}$Cs source
for temperatures between --20 and 50$^{\circ}$C. Figure~\ref{fg:resoltemp}
shows the energy resolution as a function of temperature.
The energy resolution for CLYC\_16 is 7.0\% at 20$^{\circ}$C.
The variation of energy resolution with
temperature is moderate and similar to the CsI detector of the same dimensions.

The energy resolution as a function of temperature was also measured for
CLYC\_14.
The procedures for assembling the detector were not the same as
other detectors. The results are not presented in the figure because the
comparison would not be fair. Moreover, CLYC is known to be
fragile under temperature changes and this crystal might have developed
fractures during the experiment. In spite of that, the
ability to discriminate between neutron and gamma radiation is not affected
as seen in the next section.

The energy resolution for the CLLB detector is 4.2\% at 20$^{\circ}$C and
varies very little with temperature. For temperatures below 20$^{\circ}$C,
the energy resolution improved slightly. In contrast, the energy resolution
gets worse for temperatures above 30$^{\circ}$C. Degradation of energy
resolution is observed for all the detectors studied in this work.
This phenomenon is attributed to the higher dark current in the SiPMs at
high temperatures.

The energy resolution for the NaIL detector is 7.6\% at 20$^{\circ}$C. In
contrast, the energy resolution is 6.9\% when the NaIL scintillator was
coupled to a 3-inch PMT.
The large mismatch of the contact area between the NaIL scintillator and the
SiPM array
results in worse energy resolution. For this reason, the energy
resolution for NaIL was not measured at other temperatures.


\subsection{Pulse Shape Discrimination}
\begin{figure}[!t]
\centering
\includegraphics[width=3.0in]{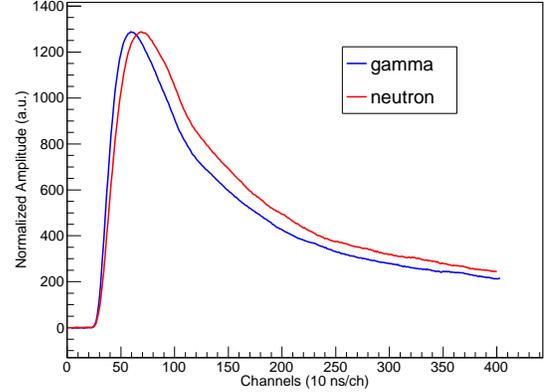}
\caption{Pulses for gamma (red curve) and neutron (blue curve) detected by CLYC
  at 20$^{\circ}$C.}
\label{fg:clycps}
\end{figure}

\begin{figure}[!t]
\centering
\includegraphics[width=3.0in]{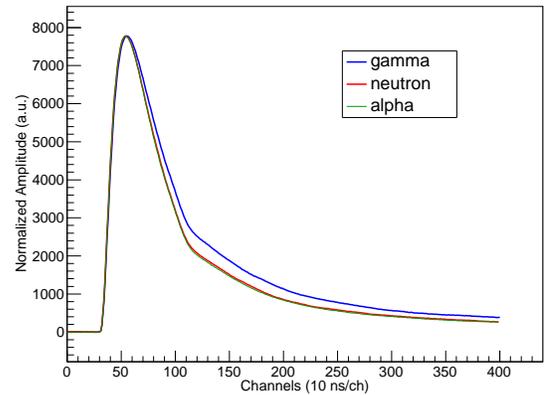}
\caption{Pulses for gamma (red curve), neutron (blue curve), and alpha (green
  curve) detected by CLLB at 20$^{\circ}$C.}
\label{fg:cllbps}
\end{figure}

\begin{figure}[!t]
\centering
\includegraphics[width=3.0in]{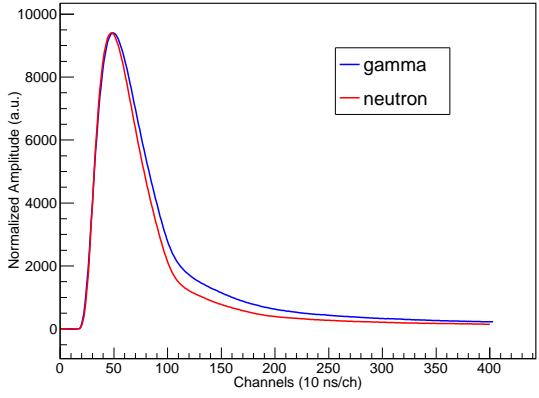}
\caption{Pulses for gamma (red curve) and neutron (blue curve) detected by NaIL
  at 20$^{\circ}$C.}
\label{fg:ps20c}
\end{figure}

Although the gamma-equivalent energy for neutrons is greater than 3~MeV for
all three scintillators, CLYC, CLLB, and NaIL, these scintillators will also
respond to high energy cosmic-rays. The fact that the pulse shape for neutrons
is different from that of gammas and cosmic background allows the use of pulse
shape discrimination (PSD) for neutron identification. While the difference in
pulse shape between neutrons and gammas is in the rise time for CLYC,
the difference in pulse shape is in the decay time for CLLB and NaIL.
Figure~\ref{fg:clycps},~\ref{fg:cllbps}, and \ref{fg:ps20c} displays the pulses
for neutrons and gammas detected by CLYC, CLLB, and NaIL, respectively, at
20$^{\circ}$C. These pulses, shown in the figures, are the average of 10 pulses
for each type of radiation and the averaged gamma pulses are normalized to the
maximum of the averaged neutron pulses.

\begin{figure}[!t]
\centering
\includegraphics[width=3.0in]{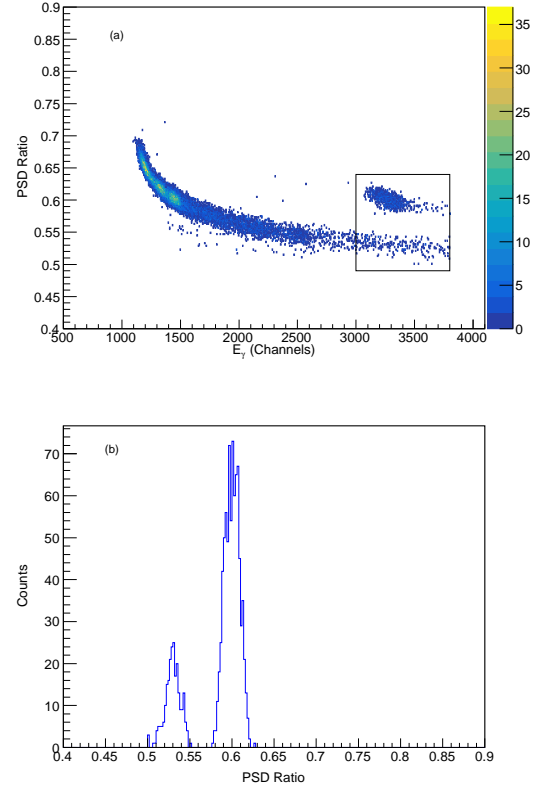}
\caption{(a) Scatter plot of PSD ratio versus gamma energy for the measurement
  with NaIL at 20$^{\circ}$. (b) Projection of
events in the rectangular area onto the PSD ratio axis.}
\label{fg:psd_20c}
\end{figure}

In this work, the PSD is accomplished by taking the ratio
of the time integral of the pulse with a short gate (Prompt) to that with
a long gate (Prompt+Delay).
\begin{equation}
PSD = \frac{Prompt}{Prompt+Delay},
\end{equation}
Figure~\ref{fg:psd_20c}(a) displays the scatter plot of the PSD ratio versus the
integral of the long gate, which is equivalent to the gamma energy, for the
measurement with NaIL at 20 $^{\circ}$C. Neutron events can be seen
as the isolated group above channel 3000. The events extending from channel
1000 to 3800 are from gammas. Two groups of gamma events between channel
1000 and 1500 are the 1.17 and 1.33~MeV gamma-rays from $^{60}$Co. Lower energy
gamma events were ignored by setting the threshold of the constant fraction
discriminator.  Figure~\ref{fg:psd_20c}(b) shows the projection of events in the
rectangular area in Fig.~\ref{fg:psd_20c}(a) onto the PSD ratio axis.

To quantify neutron and gamma discrimination, a figure of merit
(FOM) is defined as
\begin{equation}
FOM = \frac{|\mu_{\gamma}-\mu_n|}{w_{\gamma}+w_n},
\end{equation}
where $\mu$ and $w$ are the centroid and full-width-at-half-maximum,
respectively, of neutron
and gamma distributions in the histogram of PSD ratio. 

\begin{figure}[!t]
\centering
\includegraphics[width=3.0in]{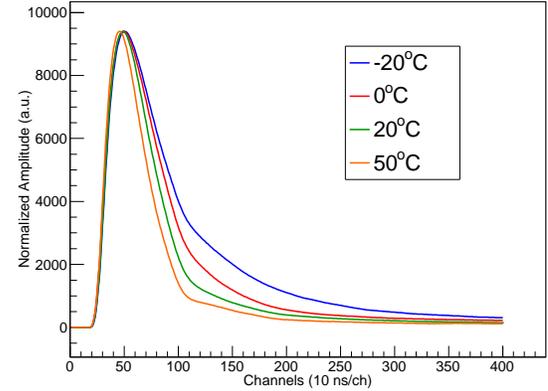}
\caption{Pulses for neutrons detected by NaIL for temperatures at
  --20 (blue curve), 0 (red curve), 20 (green curve), and 50$^{\circ}$C
  (orange curve).}
\label{fg:neutps}
\end{figure}

\begin{figure}[!t]
\centering
\includegraphics[width=3.0in]{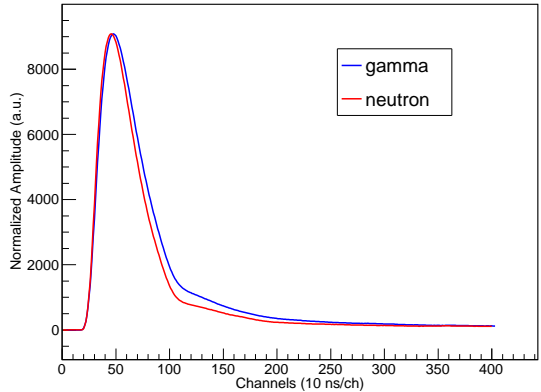}
\caption{Pulses for gamma (red curve) and neutron (blue curve) detected by
  NaIL at 50$^{\circ}$C.}
\label{fg:ps50c}
\end{figure}

The pulse shape changes with temperature for all three detectors. Shown in
Fig.~\ref{fg:neutps}
are the pulses for neutrons detected by NaIL at --20, 0, 20, and 50$^{\circ}$C.
Each pulse shown is the average of 10 pulses at the same
temperature and the amplitude of pulses at different temperatures has been
normalized to the maximum of the pulse at 20$^{\circ}$C.
Moreover, the pulse shape for neutron and gamma changes differently with
temperature. Shown in Fig.~\ref{fg:ps50c} are neutron and gamma
pulses detected by NaIL 50$^{\circ}$C. Comparing with Fig.~\ref{fg:ps20c}, it
can be seen that the difference in pulse shape between neutrons and gammas is
smaller at 50$^{\circ}$C than at 20$^{\circ}$C. Consequently, the FOM
for neutron-gamma discrimination is smaller
at 50$^{\circ}$C.

\begin{figure}[!t]
\centering
\includegraphics[width=3.0in]{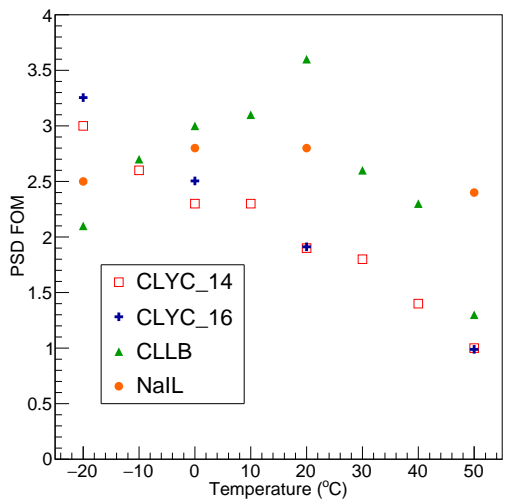}
\caption{The FOM for neutron-gamma discrimination as a function of temperature
  for CLYC\_14 (squares), CLYC\_16 (crosses), CLLB (triangles),
  and NaIL (circles).}
\label{fg:psdfom}
\end{figure}

Figure~\ref{fg:psdfom} displays the FOM for neutron-gamma discrimination versus
temperature for the scintillators studied in this work.
For CLYC, the FOM for neutron-gamma discrimination
continues to improve at low temperatures. Similar results were obtained for
a 1~cm$^3$ CLYC with a 6x6 mm$^2$ SiPM \cite{mes2016}. This behavior is not
observed for CLLB and NaIL. The higher dark current in the SiPMs
at higher temperatures contributes to the worsening of the discrimination
of neutrons and gammas by PSD. It is interesting to note that while the energy
resolution for CLYC\_14 is poor for
gamma spectroscopy measurement, this scintillator works well for
neutron-gamma discrimination. The newer CLYC sample (CLYC\_16) shows the same
FOM for neutron-gamma discrimination as CLYC\_14.

The FOM for neutron-gamma discrimination for NaIL is greater than 2 for
all the temperatures. The NaIL scintillator also has the least variation of
the FOM with temperature, as shown in Fig.~\ref{fg:psdfom}. Overall,
the FOM is greater than 1 for all three scintillators in the
temperature range studied. Therefore, the discrimination between neutron and
gamma radiation is acceptable even at high temperatures. 

\begin{figure}[!t]
\centering
\includegraphics[width=3.0in]{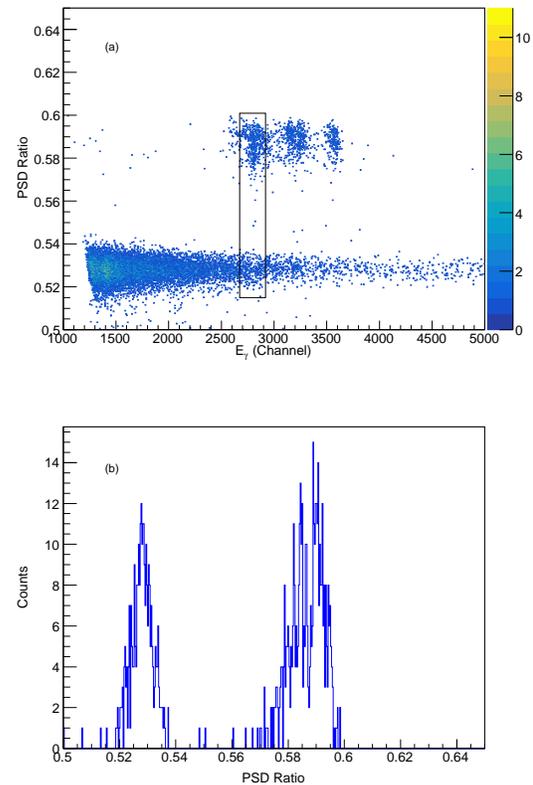}
\caption{(a) Scatter plot of PSD ratio versus gamma energy for the measurement
  with CLLB at 20$^{\circ}$. (b) Projection of
  events in the rectangular area onto the PSD ratio axis.}
\label{fg:cllbpsd}
\end{figure}

There are actinium contaminants in CLLB which emits alpha-particles. The
pulse shape for alphas and neutrons is very similar, as shown in
Fig.~\ref{fg:cllbps}. Moreover, the
gamma-equivalent energy for the alpha-particles is greater than 3~MeV.
These factors bring challenges in separating neutrons and alphas by PSD,
as shown in Fig.~\ref{fg:cllbpsd}. Because the CLLB is loaded with natural
Li, the efficiency for detecting neutrons from $^{252}$Cf is low. As a result,
the neutron peak, located between gamma and alpha peaks in
Fig.~\ref{fg:cllbpsd}(b), is barely distinguishable from
the alpha peak.
Although the FOM for neutron-gamma discrimination
for CLLB is comparable to that for CLYC and NaIL, the FOM for neutron-alpha
discrimination is less than 1 for temperatures between --20 and 50$^{\circ}$C.
In spite of
the poor neutron-alpha discrimination, the alpha background can be measured,
3.67$\pm$0.25/min, and corrected for neutron measurements.

\subsection{Neutron Detection Efficiency} 
\begin{figure}[!t]
\centering
\includegraphics[width=3.0in]{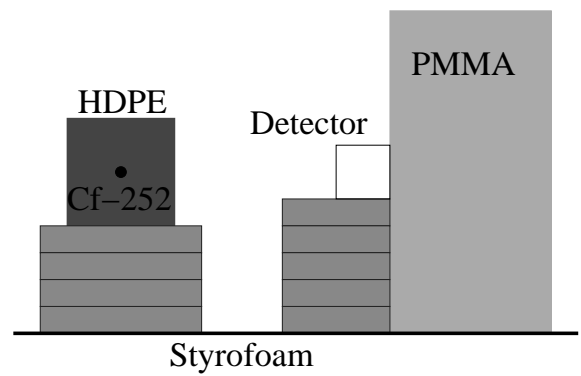}
\caption{Apparatus for measuring neutron detection efficiency.}
\label{fg:ansineut}
\end{figure}

The efficiency of neutron detection was measured for the three
scintillators and compared with a $\varnothing$15x54 mm$^3$
8~atm $^3$He tube. As shown in Fig.~\ref{fg:ansineut}, a $^{252}$Cf source was
enclosed in the center of a 10~cm diameter HDPE cylinder and placed 25~cm
from the detector. Located behind the detector is
a 15~cm thick PMMA block. Both the detector and the neutron source are
elevated by Styrofoam pads to reduce the detection of neutrons reflected by the
table surface.

The neutron detection efficiency for CLYC, CLLB, and NaIL with respect to
the 8-atm $^3$He tube is shown in Table~\ref{tb:eff}.
The CLYC and NaIL
scintillators are loaded with enriched
$^6$Li whereas the CLLB scintillator is loaded with natural Li. If enriched
$^6$Li were used in fabricating the CLLB scintillator, there would be
$2.84 \times 10^{21}$ $^6$Li/cm$^3$ compared to the 
$3.46 \times10^{21}$ $^6$Li/cm$^3$ in CLYC. Therefore, the neutron
detection efficiency for CLLB loaded with enriched $^6$Li is expected to
be close to CLYC.
\begin{table}[!t]
\centering
\caption{Neutron detection efficiency for CLYC, CLLB (loaded with natural
Li), and NaIL (1.5-inch cylinder) with
  respect to an 8-atm $^3$He tube.}
\begin{tabular}{llll} \hline
    scintillator & CLYC & CLLB & NaIL \\ \hline
    $\varepsilon/\varepsilon_{He}$ & 1.9 & 0.2 & 1.4 \\ \hline
\end{tabular}
\label{tb:eff}
\end{table}

\begin{figure}[!t]
\centering
\includegraphics[width=3.0in]{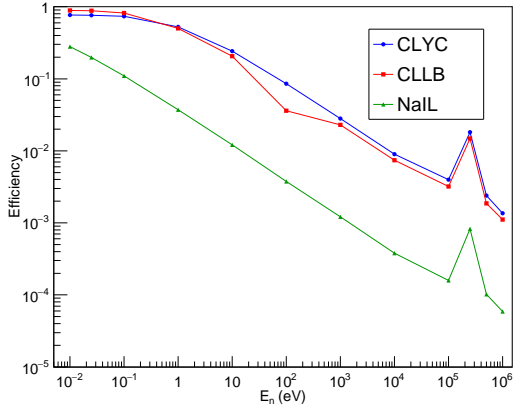}
\caption{Results of Geant4 calculations for neutron detection efficiency by
18~mm CLYC, CLLB, and NaIL. The neutron energies are between 0.01~eV and 1~MeV.}
\label{fg:deteff}
\end{figure}

Geant4 calculations have been carried out to study the neutron detection
efficiency for neutrons incident on an 18 mm of CLLB and CLYC. The neutronHP
package was used. A detection event is signified by the generation of a
triton-alpha pair in the material. Shown in Fig.~\ref{fg:deteff} are the
results of the Geant4 calculations for neutron energies from 0.01~eV to 1~MeV.
As discussed above, there is more $^6$Li per volume in CLYC, therefore, the
neutron detection efficiency is higher than CLLB.
Nevertheless, the neutron capture
by $^{35}$Cl at energies below 1~eV competes with that of $^6$Li resulting in
the detection efficiency that is lower than CLLB. Additionally,
there is a peak at 250~keV which is due to a resonance.

The NaIL scintillator has a larger volume than the CLYC and CLLB used in this
work. It is loaded with 0.5 atomic \% of $^6$Li. It has been reported that
Saint-Gobain will increase the $^6$Li concentration to 1 atomic \% for their
future NaIL products. On this account, there are $1.47\times10^{20}$
$^6$Li/cm$^3$
in NaIL which is an order of magnitude less than CLYC and CLLB.
Because the neutron absorption cross section for both Na and I is substantially
smaller than $^6$Li, the neutron detection efficiency for NaIL will not be
affected as much as CLYC at low energies.
As shown in Fig.~\ref{fg:deteff}, the neutron
detection efficiency for NaIL is only a factor of 4 less than CLYC at 0.025~eV
(thermal energy). However,
at higher energies the neutron detection efficiency for NaIL is
an order of magnitude lower than CLYC. At these energies, neutron capture by
$^6$Li competes more favorably than other elements in CLYC, therefore, the
neutron detection efficiency is commensurate with the number
density of $^6$Li. Since it is promising to
manufacture large size NaIL, this would help increase the neutron detection
efficiency \cite{sgnail}. 

In the future, it would be useful to repeat the measurement of neutron
efficiency with enriched $^6$Li loaded CLLB and an 18~mm NaIL cube.

\section{Conclusion}
Three Li-loaded inorganic scintillators, CLYC, CLLB and NaIL, have been tested
for neutron and gamma detection for temperatures between --20 and
50$^{\circ}$C. The energy resolution for CLLB is around 4\% for the 662~keV
gamma-ray which is almost a factor of 2 better than CLYC and NaIL.
Pulse-shape discrimination works well in discriminating neutrons from gammas
for all three scintillators. In particular, the FOM for neutron-gamma
discrimination for NaIL is greater than 2.0 and has the smallest variation with
temperatures. The sensitivity for detecting moderated neutrons with 
CLYC is the highest among these three scintillators and is better than an 8~atm
$^3$He tube. The neutron sensitivity
is low for CLLB because natural Li is used instead of enriched $^6$Li.
According to the experimental results, all
three scintillators are suitable for low-power and compact-geometry
applications when they are coupled to the custom SiPM array.

\section{Acknowledgment}
We would like to thank S. Lam and S. Swider of CapeSym for preparing and
providing the CLYC\_16 crystal and A. Zonneveld for loaning the NaIL crystal.

\end{document}